# Performance Comparison of MPICH and MPI4py on Raspberry Pi-3B Beowulf Cluster

*Saad Wazir, EPIC Lab, FAST-National University of Computer & Emerging Sciences, Islamabad, Pakistan.*
*E-mail: swazir.mscs18seecs@seecs.edu.pk*

*Ataul Aziz Ikram, EPIC Lab FAST-National University of Computer & Emerging Sciences, Islamabad, Pakistan.*
*E-mail: ata.ikram@nu.edu.pk*

*Hamza Ali Imran, EPIC Lab, FAST-National University of Computer & Emerging Sciences, Islambad, Pakistan.*
*E-mail: himran.mscs18seecs@seecs.edu.pk*

*Hanif Ullah, Research Scholar, Riphah International University, Islamabad, Pakistan. E-mail: hukhan.dev@gmail.com*

*Ahmed Jamal Ikram, EPIC Lab, FAST-National University of Computer & Emerging Sciences, Islamabad, Pakistan.*
*E-mail: ahmed.jamal.ikram@gmail.com*

*Maryam Ehsan, Information Technology Department, University of Gujarat, Gujarat, Pakistan.*
*E-mail: maryam.ehsan@uog.edu.pk*

**Abstract**--- Moore's Law is running out. Instead of making powerful computer by increasing number of transistor now we are moving toward Parallelism. Beowulf cluster means cluster of any Commodity hardware. Our Cluster works exactly similar to current day's supercomputers. The motivation is to create a small sized, cheap device on which students and researchers can get hands on experience. There is a master node, which interacts with user and all other nodes are slave nodes. Load is equally divided among all nodes and they send their results to master. Master combines those results and show the final output to the user. For communication between nodes we have created a network over Ethernet. We are using MPI4py, which a Python based implantation of Message Passing Interface (MPI) and MPICH which also an open source implementation of MPI and allows us to code in C, C++ and Fortran. MPI is a standard for writing codes for such clusters. We have written parallel programs of Monte Carlo's Simulation for finding value of pi and prime number generator in Python and C++ making use of MPI4py and MPICH respectively. We have also written sequential programs applying same algorithms in Python. Then we compared the time it takes to run them on cluster in parallel and in sequential on a computer having 6500 core i7 Intel processor. It is found that making use of parallelism; we were able to outperform an expensive computer which costs much more than our cluster.

**Keywords**--- MPI, Beowulf Cluster, SBC, HPC, MPI.

## I. Introduction

There is an always and increasing demand of cheap, small and computationally powerful computers around the globe. Moore's Law is running out. One way to fulfill the demand of more computational power is parallel computing. Instead of making more powerful computer by increasing number of transistors now we are moving toward Parallelism. There is an increasing demand of computationally powerful computers. Areas requiring great computational speed include numerical simulation of scientific and engineering problems. Such problems often need huge quantities of repetitive calculations on large amount of data to obtain valid results. Few examples are, modeling of DNA structures, global forecasting of weather and etc. Moreover, IOTs (Internet of Things) are generating a lot of data daily and that data is to be processed to extract some useful information from it.

A computer cluster is a collection of cooperating computers. There are several variations of these, one of them is called Beowulf Cluster. A Beowulf cluster is a uniform collection of off the shelf computers connected by an Ethernet network. An important distinguishing feature is that only one computer, the head node, is communicating with the outside network. A Beowulf cluster is dedicated only to jobs assigned through its head node. Parallel programming differs from traditional sequential programming. Additional complexity becomes apparent when one must coordinate different concurrent tasks. Super Computers such as IBM Roadrunner or Cray Jaguar having many processors in them or Beowulf cluster which are actually cluster of cheap commodity hardware. They require a network and few libraries installed on them which help them divide their task or problems. But the problem with both of these solution is they are highly expensive and a huge in size and demand large place to station them. The history of Beowulf cluster was started at NASA when they needed a supercomputer for data analysis, but couldn't buy one of sufficient power for a cheap enough price. It was built out of 16 personal computers that were 66 MHz, 486 processors, with standard off-the-shelf network cards, and the open source operating system Linux.

The Raspberry-Pi is like an SBC capable of many functionalities similar to super-computing while being portable, economical and flexible. Running on open source Linux, makes it a preferred choice for acquiring more user base for research purposes. It also allows for creating a low-cost, low energy, portable cluster that attends to the needs of small or medium sized users. Generally, one R-pi would be assigned the task of the main server or as mentioned above the head node, and required number of nodes added to it to form the cluster. One of the advantages of such clusters is the direct control over each node.

This also benefits that failure of a node does not affect the computations running through the rest of the cluster. Therefore, nodes can be independently programmed to interact with each other, in addition to receiving and transmitting data and instructions from and to the server. The basic mechanism that enables implementing such clusters is the 'parallel processing' capability. Research enthusiasts now can also use the embedded networking capability to design portable clusters that replace the costlier machines. Also by using mainstream tools such as MPI and Hadoop on such clusters, cloud computing capabilities can also be devised and deployed. This paper is not about going in depth into parallel programming, but the focus lies more on how to design and build a cluster to be used more likely as a supercomputer.

## II. Literature Review

The application of SBC to parallel processing and cluster computing has been a topic of wide interest since the past few years. Anil Jaiswal [16] presented a study of different benchmarking clusters for the programming of Raspberry pi clusters considering the popular parallel programming architectures such as openMP, MPI and MapReduce framework. A little expensive, but one of the pioneers in cluster assembling was developed by Cox et al. [1] in 2012 at the University of Southampton for a total cost of around £3400 a cluster, called Iridis-pi, of 64 Raspberry Pi nodes with 256 MB main memory per node. The nodes were powered by using 64 individual 5 V power supplies which was the main hurdle. Also in 2012 at University of Edinburgh, Balakrishnan [2] constructed a cluster by using six PandaBoard SBC and two Raspberry Pi nodes, which performed 6.484 Gflops using the six PandaBoard nodes.

The addition in the research was the use of multiple SBCs. A cluster of 32 Raspberry nodes with 512 MB main memory per node, was devised by Kiepert in 2013 [3][4], in which two power supplies were powering the complete system utilizing a total power usage of 167 W. A cluster called MegaRPi was presented by Abrahamsson et al. [5] in 2013 a cluster, consisting of 300 Raspberry Pi nodes with 512 MB main memory per node with standard PC power supplies. The research presented a comparison between the power consumption of a single node with other computers while running a standalone HTTP server. Unfortunately, no further power measurements or Gflops results were presented. Sukaridhoto et al. [6] presented a cluster of 16 Pandaboard nodes which using a single 200 W, 5 V, 40 A power supply to power the nodes but the power consumption of the entire cluster was not presented. Ou et al. [7] compared the performance, energy-efficiency and cost-efficiency of a single PandaBoard computer with an Intel X86 workstation for the three applications which were web server throughput, in-memory database and video transcoding.

The work examined the number of nodes a cluster required to compete with the workstation. Tso et al. [8] presented a distinctive Raspberry Pi Cloud modelled as a data center, composed of 56 Raspberry Pi Model B nodes. The research presented the comparison analysis of the acquisition cost, electricity costs and cooling requirements of the SBC cluster with a testbed of 56 hardware servers without highlighting any performance measurements. Pfalzgraf and Driscoll [9] used a single 600 W power supply to power cluster of 25 Raspberry Pi nodes. It does not provide any power or performance results. An important study for multitask computing using Python was presented by Monte Lunacek et al [15] using mpi4py as a baseline for the comparisons of IPython parallel and Celery for scaling study using over 12000 cores and multiple thousand tasks. An evaluation of python based libraries was presented by Jacob Morra and Qusay Mahmoud [14] for the run time computation of efficiency for distributed tasking including PyRO v4.45, DCM v1.0.0, PP v 1.6.4.4 and Mpi4py v2.0.0.0. In 2016 Abrrachman Mappuji et al. [11] presented a utilization of multiple Raspberry pi 2 SBCs as a cluster to compensate its computing power. The findings were that the increase in every SBC member in a cluster is not necessarily a measure for a significant increase in computational capabilities, and also recommending that 4 nodes are a maximum for an SBC cluster based for optimum power consumption.

Christian Baun [12] presented a similar study for building inexpensive cluster systems of SBCs and analyzing their energy efficiency with the High Performance Linpack (HPL) benchmark. Also taking energy efficiency as a challenge and generating a small scale data center by the use of SBCs was researched by Basit Qureshi and Anis Koubaa [13], with the construction of two clusters having 20 nodes each. The findings showed that although the low cost of building a cluster is often looked-for, the clusters do yield low energy efficiency due to limited onboard capabilities.

These works have shown the potential of clusters of SBC, but little work is notable to have implemented any algorithm on the system further giving a comparison analysis afterwards. Python is becoming famous in the field of high performance computing. The reason of its popularity is its ease in programming and its big community. This paper is focused on the evaluation that how much one is ready to lose for getting ease in programming. For that purpose, a comparison of MPICH and MPI4py was performed. MPICH is one of the most famous open source implementation of MPI; and is now becoming a standard for supercomputing.

Moreover, MP4py is a python binding for MPI and is the one of the best Python tools available for distributed computing (one of the papers prove it). As the true performance evaluation factor is a time taken by program to run an empirical analysis for performance comparison was evaluated. For that a cluster of Raspbery pi 3B has been developed. The reason for using them was they were one of the cheapest and reliable method to create a cluster that resembles current time Supercomputers. Monte Carlos simulation for finding the value of pi was the method selected to be performed. Then the time taken by the program running different processes was measured. The code was written in C using MPICH and complied using mpicc. Along with that, the same algorithm was implemented in python 2 using MPi4py and compered the time taken for the same tasks.

### III. Methodology And Implementation

Following the aforementioned purpose of the research, an SBC cluster is developed. Due to the availability of a better and improved latest hardware i.e Raspberry-pi 3, a better cluster has been anticipated and thus implemented which is explained in the following sections.

#### A. Hardware Selection

After exploring multiple boards including Parallela, Pine64, Pine64+, Raspberry pi 3B, Asus Tinker Board, Raspberry pi zero, the short listed boards are Pine64+ (1GB edition) and Raspberry pi 3B. There hardware specifications are in the figure 1 below:

| Pine64+ | Raspberry Pi 3B |
|---|---|
| • Cost (original does not include shipping cost): 20$ for 1GB Edition | • Cost (original does not include shipping cost): 37.8$ |
| • Processor: 1.2 GHZ quad-core ARM Cortex A53 (ARMv8 Instruction Set) | • Processor: 1.2 GHZ quad-core ARM Cortex A53 (ARMv8 Instruction Set) |
| • Dual Core Mali 400 MP2 Graphics card | • GPU: : Broadcom VideoCore IV @ 400 MHz |
| • Ethernet: 10/100/1000 Mbps Ethernet Port | • Ethernet: 10/100 Mbps Ethernet Port |
| • RAM:1 GB DDR3 | • RAM:1 GB DDR2 |

Figure 1: Specifications of Shortlisted SBC

Pine64+ has better hardware and cost far less but it has very poor software support and does not have any stable Linux distribution with a limited developer community. While Raspberry pi has a large user base and amiable software support. Therefore, the obvious candidate for selection is Raspberry pi 3B. TP-Link 8 port 10/100 Switch was used for network creation due to an adequate cost.

### B. Software Selection

Software Selection includes two things libraries selection and Operation System selection. Out of the available open source libraries like OpenMpi, MPICH, MPI4py, IBM MPI, the chosen two libraries are MPICH and MPI4py which allows the users to program in Python and C++ respectively.

There are many stable Linux distributions available for Raspberry pi like Raspbian, Ubantu Mate, Arch Linux etc. The obvious candidate is Raspbian because it is the official Operating System by Raspberry Pi Foundation's latest version "Raspbian Stretch".

### C. Block and System Level Diagrams

The block Diagram of the cluster is given below as Figure 2. The System has one master node and all other nodes act as slave nodes as can be seen in the Block Diagram. For communication of messages between nodes a network is needed which is further created with a second layer switch and CAT-5 cables.

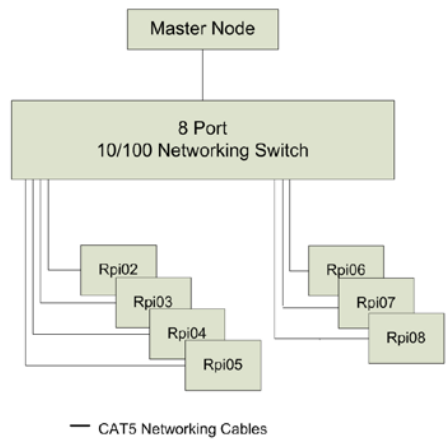

Figure 2: Block Diagram

Similarly, the system level diagram is shown in Figure 3.

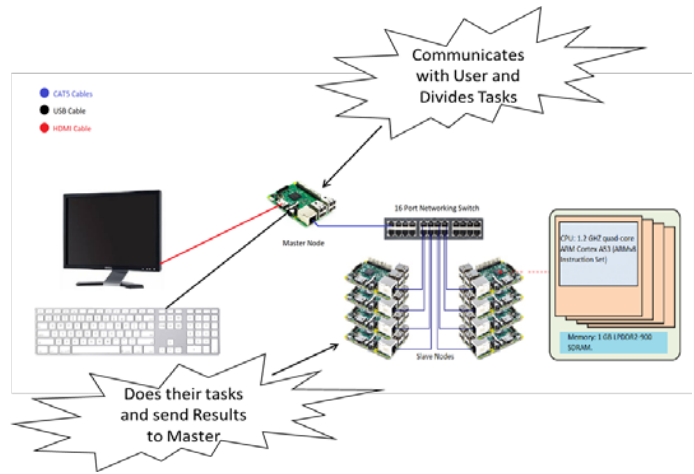

Figure 3: System Diagram

### D. Working

Computing devices in general comprise of multiple conditions information is exchanged between different processes. General purpose computers utilize communications through shared memory, in which two or more processes read and write to a shared section of memory [10] which is unreliable because we have both CPU and memory elements in each cluster member. In this case passing of the information or messaging is basically through packets of information in predefined formats moving between processes by the manager called interface. The standard MPI is used because standardization, portability, performance opportunities, functionality and availability.

MPI is a worldwide accepted standard for supercomputers' communication. When any program has been run which uses MPI, the number of processes to create needs to be specified. Each process is actually the same copy of the program but only certain part of the code is being executed by each process. The selection of the portion is being done on the basis of a unique ID which is being assigned by MPI to each process and it is called Rank. MPI allows different processes to send and receive messages between them. A communicator is a group of Processes that can communicate. By default, MPI creates a communicator named MPI_COMM_WORLD. This communicator has all the processes being created while executing the program as shown in figure 4 below. Communicators can be created by using tools provided by MPI. Simple programs typically only MPI_COMM_WORLD is needed.

The following Flowchart (Figure 4) explains who any processes created at the time of execution of the program works.

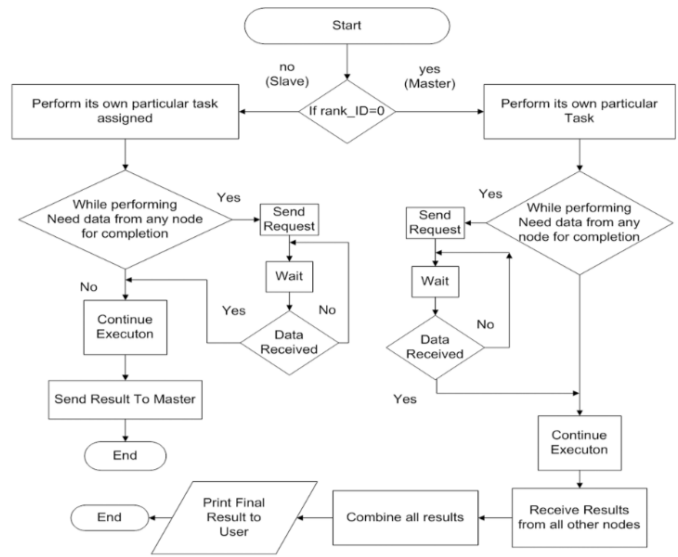

Figure 4: Flowchart

### E. Algorithms Implemented

The objective of taking this research was to create a cheap affordable replica of current time supercomputer and to also learn how to work on them. For that purpose, two algorithms have been implemented in two most famous implementations of MPI, namely MPI4py and MPICH.

### 1) Monte Carlo's Method

Suppose we have a square of Area 2x2 and then further assume we have placed a circle of radius 1 exactly at its center. Now if we take the Ratio of areas of both of these figures we will get:

$$\frac{Area\ of\ circle}{Area\ of\ Square} = \frac{\pi(1)^2}{2x2} = \frac{\pi}{4}$$

Now if we are somehow able to find out the ratio and multiply that by 4 we will be value of Pi. In our programs we are only considering the first quadrant of the figure as shown in figure 5. Then we are generating two random numbers between 0 and 1 and making use of equation of circle and checking that whether the radius is 1 or less than 1 means inside the circle or not if it is then it is a hit. The ratio is being calculated by taking the ratio of hits over total tries. The following flowchart explains the working of parallel program (figure 6).

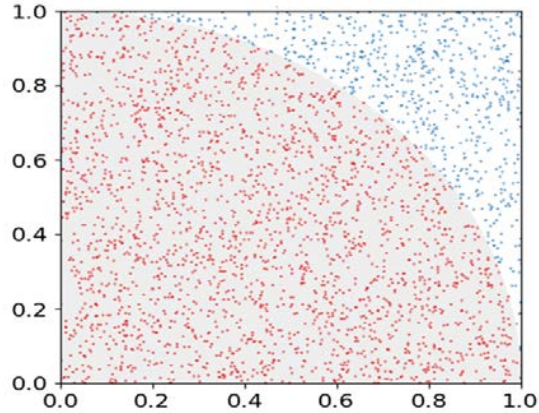

Figure 5: First Quadrant

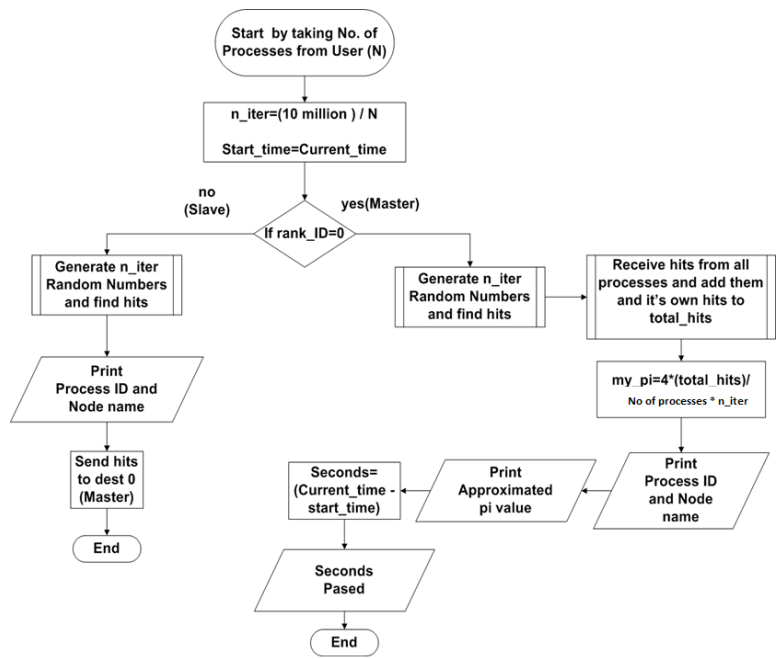

Figure 6: Parallel Algorithm of Monte Carlo Method

### 2) Prime Number Generator

To extract parallelism for prime number generation we actually divided the range (which in our case was 0 to 20,000) into sub ranges. So each process was then finding prime numbers in its own range sequentially. To understand this algorithm, consider the following example. Suppose the range is 0 to 1000 and we have created 4 processes. To the division of ranges is done by following method:

$$(Rank * N, (Rank + 1) * N)$$

Where,

- Rank is the ID of process
- N is the number of processes.

So for our case division will be as follows figure 7:

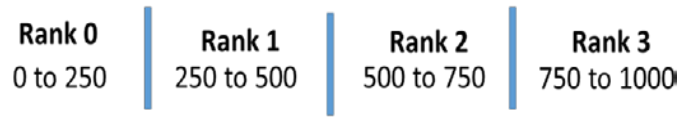

Figure 7: Prime Number Generation Ranks

The flowchart explaining its parallel program is given below figure 8:

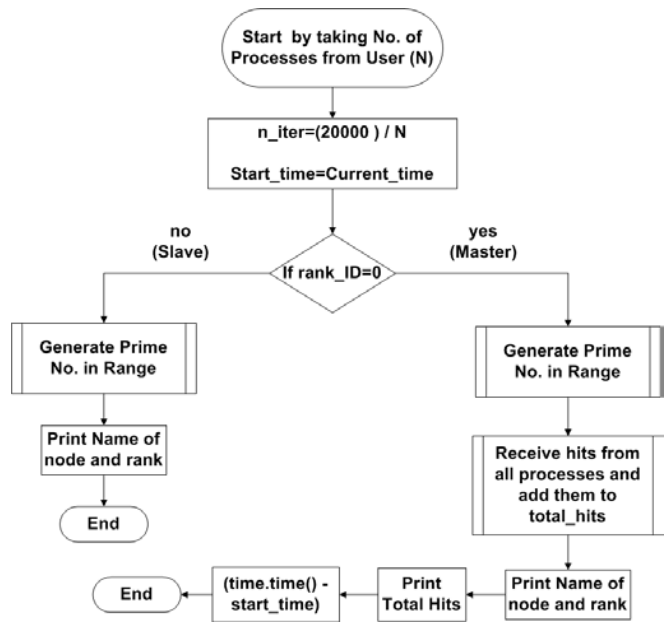

Figure 7: Flowchart of Prime Number Generator

## IV. Results and Discussions

We have compared the time taken by parallel program written in Python and C++ running on cluster (by creating different number of processes) with sequential program running on 6500 core i7 Intel processor (on single core). The results of both simulations are given below in figure 8, (9 and 10) respectively.

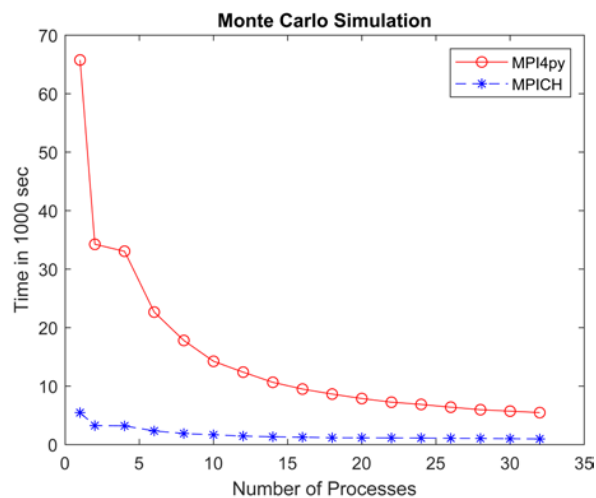

Figure 8: Monte Carlo's Simulation Time Graph

The time difference between the execution time of Prime number generation for MPI4py and MPICH was too much that plotting their results on same graph in a nice manner was not possible we have made two graphs for each of them. Figure 8 shows graph for MPI4py and figure 9 shows for MPICH.

The results showed above shows us that parallelism does improve performance but up to a certain level. After that increase in number of processes does not give us much improvement in performance with that performance obtained by C++ in parallel is about 10 times more than Python. Which leads us to a conclusion that although writing code in python is far less time consuming and is easy it still better to spend more time on coding in C++ .Because the difference in performance is too much.

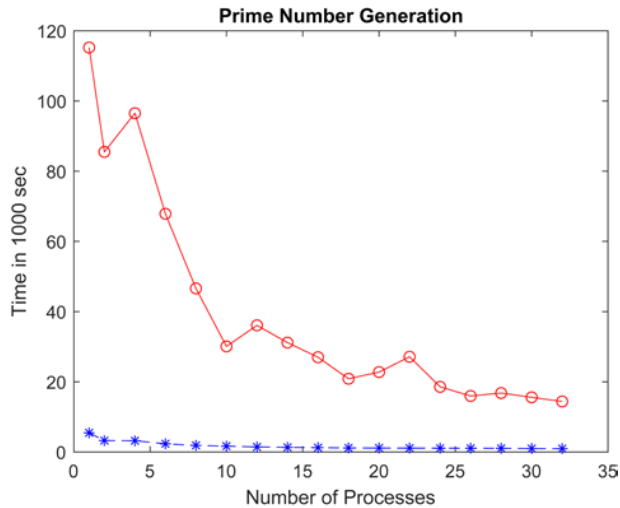

Figure 9: Prime No with MPI4py

## V. Recommendation and Future Work

We have written parallel programs of Monte Carlo's Simulation for finding value of pi and prime number generator in Python and C++ making use of MPI4py and MPICH respectively. We have also written sequential programs applying same algorithms in Python. Both are then compared with the time it takes to run them on cluster in parallel and in sequential on a personal computer having 6500 core i7 Intel processor. It is found that making use of parallelism, we were able to outperform an expensive computer which costs much more than our cluster.

Because the parallelism of a multi-core computer, SBC is needed to optimize the computing performance. It is better to test the cluster with parallel optimized high performance computing benchmarking programs with utilization of concurrent programming concepts in the future. Utilization of another kind of operating system is needed to enrich the knowledge of SBC cluster computer in the future work, or better off, to devise a multiplatform user interface for the cluster. Implementation of different algorithms on the cluster can also be an area of interest in the future to test the robustness of the system.